\documentclass{iopart}
\usepackage{iopams}

\usepackage[english]{babel}

\usepackage{amsfonts}
\usepackage{amssymb}
\usepackage{amscd}
\usepackage{epsfig}

\usepackage[latin1]{inputenc}

\def\sh{\sinh}

\def\a{\alpha}
\def\b{\beta}
\def\d{\delta}

\def\k{\kappa}

\newcommand{\csh}{\mbox{\rm cosech}}

\newcommand{\Sc}{Schr\"odinger }
\newcommand{\eqref}[1]{(\ref{#1})}

\newcommand{\be}[1]{\begin{equation}\label{#1}}
\newcommand{\ee}{\end{equation}}
\newcommand{\ba}[1]{\begin{eqnarray}\label{#1}}
\newcommand{\ea}{\end{eqnarray}}

\newcommand{\diag}{\mbox{\rm diag}}

\begin{document}
\title[Coupling between scattering channels with SUSY transformations]%
{Coupling between scattering channels with SUSY transformations
for equal thresholds}

\centerline{\today}

\author{Andrey M Pupasov\footnote{Boursier de l'ULB}$^{1, 2}$,
 Boris F Samsonov$^1$, Jean-Marc Sparenberg$^2$
and Daniel Baye$^2$}

\address{$^1$ Physics Department, Tomsk State University,
36 Lenin Avenue, 634050 Tomsk, Russia}

\address{$^2$ Physique Quantique,
 C.P.\ 229, Universit\'e Libre de Bruxelles,
B 1050 Bruxelles, Belgium}

\eads{\mailto{pupasov@phys.tsu.ru}, \mailto{samsonov@phys.tsu.ru},
 \mailto{jmspar@ulb.ac.be}, \mailto{dbaye@ulb.ac.be}}

\begin{abstract}
Supersymmetric (SUSY) transformations
of the multi-channel  Schr\"odinger equation with equal thresholds and
arbitrary partial waves in all channels are studied.
The structures of the transformation function and the superpotential
are analyzed. Relations between Jost and scattering matrices
of superpartner potentials are obtained.
In particular, we show that a special type of SUSY transformation
allows us to introduce a coupling between scattering channels starting
from a potential with an uncoupled scattering matrix.
The possibility for this coupling to be trivial is discussed.
We show that the transformation introduces bound and virtual states
with a definite degeneracy at the factorization energy.
A detailed study of the potential and scattering matrices is given
for the $2\times 2$ case.
The possibility of inverting coupled-channel scattering data
by such a SUSY transformation is demonstrated by several examples
($s-s$, $s-p$ and $s-d$ partial waves).

\end{abstract}

\section{Introduction}

The inverse problem in quantum scattering theory plays an
essential role in theoretical and
mathematical physics.
Following the needs of soliton theory \cite{ST1,ST2} and elastic
scattering processes in atomic and nuclear collisions
\cite{newton:82,taylor:72},
the one-channel inverse problem reached a high
perfection level of development summarized in a number of books
(see e.g.\ \cite{IS1,IS2}).
The most promising way of solving this problem
 from a pragmatic viewpoint is related with the
so called method of Darboux transformations \cite{Matveev-Salle}
which is also known under the name of supersymmetric (SUSY) quantum
mechanics \cite{SQM1,SQM2} (for a review of inverse scattering
with SUSY quantum mechanics, see \cite{BSpJPA}). In contrast, the
multichannel inverse scattering theory, which describes inelastic
scattering processes \cite{newton:82,taylor:72}, only gained a
fragmental level of development (see e.g.\ \cite{IS2,BSpJPA}).
In particular, after a few papers devoted to an application
of SUSY transformations to multichannel inversion published by
Cannata et al between 1988 and 1992
\cite{amado:88b,amado:88a,amado:90,cannata:92,cannata:93} and two
papers published in 1997 \cite{sparenberg:97} and
2000 \cite{Leeb}, where phase equivalence is discussed,
a publication gap occurred till 2006. We think that the main reason
for this inactivity is the lack of a detailed analysis of properties of
 general SUSY transformations applied to the scattering
 inversion in the multichannel case.

For potentials with different thresholds such an analysis
permitted us to introduce non-conservative SUSY
transformations \cite{sparenberg:06},
with the help of which, starting from a decoupled potential,
new exactly solvable coupled potentials were obtained
\cite{samsonov:07,pupasov:08a,sparenberg:08}.
We were also able to give a careful
analysis of the scattering properties of
the known Cox potential \cite{cox:64}
%\cite{quant-ph}
and apply it to construct a realistic model
of the magnetic Feshbach resonance phenomenon
\cite{pupasov:08b}.
For equal thresholds, to the best of our knowledge,
only two papers exist \cite{amado:88b,cannata:92},
where only a particular type of SUSY transformation is studied,
i.e.\ a transformation removing a bound state.

In the current paper, with the equal threshold inverse problem in mind,
we rather concentrate on the necessary ingredients
for a single SUSY transformation to generate coupled scattering matrices,
starting from a decoupled potential.
We first show that such a transformation introduces
bound and virtual states
at the same energy
and we calculate their degeneracy.
Next, we discuss the possibility to get a trivially
or non trivially coupled scattering matrix when both the potential and Jost matrix are non trivially coupled.
By trivially coupled scattering and Jost matrices,
we mean non-diagonal matrices which may be diagonalized by
energy-independent transformations.
Similarly, a multichannel potential is nontrivially coupled
if its matrix cannot be diagonalized by an $r$-independent
transformation, where $r$ is the radial coordinate.
Below, we will be able to answer the following questions:
does a non trivial coupling of the potential imply a non trivial coupling of the scattering matrix?
Does a non trivial coupling of the Jost matrix imply a non trivial coupling of the scattering matrix?

The paper is organized as follows.
 In the next section, we fix our notations
regarding
SUSY transformations as applied to
multichannel scattering theory.
In Section \ref{sec:MT}, we analyze
the asymptotic behaviour of the superpotential for
SUSY transformed potentials and establish conditions to get potentials with non-trivially coupled $S$-matrices from uncoupled potentials.
The analysis of the Jost-matrix determinant permits us to
find modifications produced in the spectrum by SUSY
transformations.
The rest of the paper is devoted to a detailed
analysis of the most important case from a practical point of
view, i.e.\ the case of two-channel systems which can model a rich
class of physical systems
 \cite{newton:57}.
In Section \ref{sec:LR}, the long range behaviour of the transformed
potential is studied.
In particular, we indicate an unusual
possibility for a SUSY transformation consisting in the
interchange of the centrifugal terms by the scattering channels.
In
section \ref{sec:ST}, we find the Jost and
scattering matrices, phase shifts and mixing angle for the
SUSY partner potential.
Section
\ref{sec:E} first presents examples
 of coupled potentials with
trivially coupled scattering matrices.
Then we exemplify
coupled potentials
 ($s-s$, $s-p$ and $s-d$ partial waves)
with non-trivially coupled scattering
matrices and analyze their eigenphase shifts.
In the Conclusion, a summary of the obtained results is given.

\section{Outline of scattering theory and SUSY quantum mechanics}

Our starting point is the $N\times N$ multi-channel \Sc equation ($E=k^2$)
\begin{equation}\label{se1}\fl
H\psi(k,r)=k^2\psi(k,r)\,,\qquad
H=-I_N\frac{d^2}{dr^2}+V(r)\,,\qquad
V=\frac{l(l+1)}{r^2}+\bar{V}(r)\,,
\end{equation}
where $r$ is the radial coordinate, $r\in(0,\infty)$, $\bar{V}(r)$
is an $N\times N$ real short ranged symmetric potential matrix,
$l=\diag(l_1,\ldots,l_N)$, $l_j=0,1,\ldots$, is a diagonal matrix
of angular momentum quantum numbers, and $\psi$ may be either a
matrix-valued or a vector-valued function. Here and in what
follows we also denote $l\pm1$ (and $\nu\pm1$ below) matrix $l\pm
I_N$ ($\nu\pm I_N$) where $I_N$ is the $N\times N$ identity matrix.
Matrix $\nu$ determines the singularity of the potential at the
origin
\begin{equation}\label{Vzero}
\fl V(r\rightarrow 0)=r^{-2}\nu(\nu+1)+{\rm O}(1)\,,\quad \nu={\rm
diag}\left[\nu_1,\nu_2,\ldots,\nu_N\right], \quad \nu_j=0,
1,\ldots\,.
\end{equation}
Note that
$V$ does not contain a Coulomb-like $r^{-1}$ singularity.
Matrix $l$ defines the asymptotic behaviour of the potential
at large distances
\begin{equation}\label{Vinf}
V(r\rightarrow \infty)= r^{-2}l(l+1)+{\rm o}(r^{-2})
\end{equation}
which is typical of coupled channels involving various partial waves.

We define the regular $N \times N$ matrix solution $\varphi(k,r)$ of
\eqref{se1} according to \cite{newton:82}
\begin{eqnarray}\label{regs}
\varphi(k,r\rightarrow 0)&\to&
%{\diag}(r^{\nu_1+1},\ldots,r^{\nu_N+1})
%\left({\rm diag}[(2\nu_1+1)!!,\ldots,(2\nu_N+1)!!]\right)^{-1}
{\diag}(r^{\nu_1+1}/(2\nu_1+1)!!,\ldots,r^{\nu_N+1}/(2\nu_N+1)!!)
\nonumber \\ &=:&r^{\nu+1}[(2\nu+1)!!]^{-1}\,.
\end{eqnarray}
The Jost solution $f(k,r)$ has the usual exponential asymptotic
 behaviour
at large distances
\begin{equation}\label{jsas}
f(k,r\rightarrow\infty)\to
{\rm diag}[e^{ik_1r},\ldots,e^{ik_Nr}]:=
e^{ikr}\,.
\end{equation}
The Jost solutions $f(k,r)$ and $f(-k,r)$ form a basis in the
matrix solution
space.
The regular solution is expressed in terms of the Jost
solutions as
\begin{equation}\label{rajs}
\varphi(k,r)=\frac{i}{2k}\left[f(-k,r)F(k)-f(k,r)F(-k)\right],
\label{FdefI}
\end{equation}
where $F(k)$ is known as the Jost matrix.
From \eqref{regs} and \eqref{rajs} it follows that
\begin{equation}\label{fd1}
F(k)=\lim\limits_{r\rightarrow 0}
\left[f^T(k,r)r^\nu \right][(2\nu-1)!!]^{-1}\,.
\end{equation}

The physical solution, which appears in the partial wave decomposition
of the stationary scattering state, behaves as
\begin{equation}
\Psi(k,r\rightarrow\infty)\propto e^{il\frac{\pi}{2}-ikr}-
e^{-il\frac{\pi}{2}+ikr}S(k)\,. \label{p}
\end{equation}
The scattering matrix $S(k)$ is expressed in terms of the Jost
matrix as
\begin{equation}\label{SdefI}
S(k)=e^{il\frac{\pi}{2}}F(-k)F^{-1}(k)e^{il\frac{\pi}{2}}\,.
\end{equation}
Below we will consider SUSY partners $\tilde{V}(r)$ of the potential
$V(r)$
and find the corresponding
partners for both the Jost and scattering matrices.

For the reader's convenience,
we now describe the general scheme of SUSY transformations.
It is known that solutions of the initial \Sc equation
\eqref{se1} may be
mapped into solutions of the transformed equation
with help of the differential-matrix operator $L$,
\begin{equation}\label{psit}
\tilde{\psi}(k,r)=
L\psi(k,r):=\left[w(r)-I_N\frac{d}{dr}\right]\psi(k,r)\,.
\end{equation}
The transformed \Sc equation has form \eqref{se1} with a new potential
\begin{equation}\label{Vt}
\tilde{V}(r)=V(r)-2 w'(r)\,,
\end{equation}
Here the prime denotes the derivative with respect to $r$.
Matrix $w(r)$, called the superpotential,
is expressed in terms of a matrix solution $u$ of the initial
\Sc equation
\begin{equation}\label{shrtr}
H u(r) = -\kappa^2 u(r)\,,
\end{equation}
as follows
\begin{equation}\label{U}
w(r)=u'(r) u^{-1}(r)\,.
\end{equation}
Here $\kappa$ is
known as the factorization wave number. It
corresponds to an energy ${E}=-\kappa^2$,
called the factorization energy. Solution $u(r)$ is called
the factorization solution or
(matrix-valued) transformation function.

Let us also mention that, since $Lu(r)=0$, solution
$\tilde{\psi}(i\k ,r)$ of the transformed equation corresponding
to the energy ${E}=-\kappa^2$ is found as matrix
$\Phi(r)=[u^\dagger(r)]^{-1}$. Moreover, this matrix, when chosen
as the transformation function for the next transformation step,
cancels the previously produced potential difference. This
means that it corresponds to a transformation in the opposite
direction.

In the next section we introduce a transformation which
mixes
scattering channels in an uncoupled potential
\begin{equation}\label{vdi}
V(r)=V_d(r):= {\rm
diag}\left(V_{d;1},V_{d;2},\ldots,V_{d;N}\right)
\end{equation}
with given values of $\nu$ and $l$.
Such potentials
 may be obtained, for instance, by
 proper chains of usual 1-channel transformations
 of the zero initial potential
 (see e.g.\  \cite{samsonov:03}) resulting in
 generalized Bargmann potentials
(for more details see e.g.\ \cite{samsonov:95}).

\section{Coupling transformations \label{sec:MT}}

In the most general case the transformation
function may be expressed in terms of the Jost solutions as follows
\begin{equation} \label{tf2}
u_c(r)=f_d(-i\k ,r)C+f_d(i\k ,r)D\,,
\end{equation}
where the real constant matrices $C$ and $D$ are arbitrary.
Here and in what follows subscripts $d$ and $c$
stand for quantities related with diagonal
(uncoupled)
and non-diagonal (coupled)
 matrices, respectively.
To obtain a Hermitian potential after a transformation
with transformation function \eqref{tf2}
matrix $C^TD$ should be symmetric \cite{samsonov:07}
\begin{equation}\label{restr}
D^TC=C^TD\,.
\end{equation}
We need the asymptotic behaviour of the superpotential
$w_{c\infty}:=\lim\limits_{r\rightarrow\infty} w_c(r)$ to find the
transformed Jost solution and, hence, the Jost and scattering
matrices. As it was shown in \cite{samsonov:07} for different
thresholds, this behaviour of the superpotential depends
crucially on matrix $C$. Below we shortly discuss the method
developed in \cite{samsonov:07} while making necessary changes to
adjust it for the case of equal thresholds.

Matrices $C$ and $D$
with a maximal number of independent parameters
guaranteeing the Hermitian character of the
superpotential \eqref{U}
 have the following canonical form,
\begin{equation}\label{C12}
C=\left(
\begin{array}{cc}I_{M}&0\\ Q &0\end{array}\right),
\qquad
D=\left(
\begin{array}{cc}X_0&-Q^T\\ 0 &I_{N-M}\end{array}\right),
\end{equation}
where $X_0=X_0^T$ is a real symmetric nonsingular $M\times M$ matrix,
and $Q$ is an $(N-M)\times M$ real matrix so that
 ${\rm rank\,}C=M$.

The asymptotic matrix
$w_{c\infty}$ is determined by the behaviour of
transformation function \eqref{tf2} at large distances
\begin{equation}\label{ucas}
\fl u_c(r\rightarrow\infty)\to  A
\left(
\begin{array}{cc} I_Me^{\k r}&0\\
0 &I_{N-M}e^{-\k r}\end{array}\right),\qquad
A=\left(
\begin{array}{cc}I_M &-Q^T\\ Q &I_{N-M}\end{array}\right)\,.
\end{equation}
From (\ref{U}) and \eqref{ucas} we obtain
\begin{equation}\label{asi}
w_{c\infty}=
 \k A \left(
\begin{array}{cc} I_M& 0\\ 0 & -
I_{N-M}\end{array}\right)A^{-1}
\end{equation}
with
\begin{equation}
 A^{-1}=A^T\left(\begin{array}{cc}I_M+Q^TQ&0\\
0 &I_{N-M}+QQ^T\end{array}\right)^{-1}\,.
\end{equation}

Comparing this result with that obtained in
 \cite{samsonov:07}, we conclude that
the main difference between equal and different thresholds is the
non-diagonal character of the superpotential at infinity. Note
that superpotential $w_{c\infty}$ has a richer structure than that
previously reported by Amado et al \cite{amado:88b}. Their result
corresponds to the choice
 $M=1$ when $w_{c\infty}$
is expressed in terms of a single $(N-1)$-vector
$Q=(q_1,\ldots,q_{N-1})^T$.

Once $w_{c\infty}$ is determined one can calculate
the Jost solution $f_c(k,r)$ and
the Jost matrix
$F_c(k)$
 for the transformed potential $V_c$.
The Jost solution takes the form
\begin{equation}\label{jsc}
f_c(k,r)=L_cf_d(k,r)(w_{c\infty}-ikI_N)^{-1}\,,\quad L_c:=w_c(r)-I_N\frac{d}{dr}\,.
\end{equation}
The factor
$(w_{c\infty}-ikI_N)^{-1}$ is introduced to guarantee the correct
asymptotic behaviour of $f_c(k,r)$ (see \eqref{jsas}).
In order to find the Jost matrix, we first consider the behaviour of the superpotential
in a vicinity of $r=0$
which depends on the character of the transformation
solution \eqref{tf2}. Below we will assume that there is
no bound state at the factorization energy,
${\rm det}F_d(i\k )\neq 0$, and each column of the transformation solution
is singular at the origin.
Using the property
\begin{equation}\label{zb}
f_d(-k,r\rightarrow 0)=f_d(k,r\rightarrow 0)F_d(-k)F_d^{-1}(k)+{\rm o}(r^{\nu})\,
\end{equation}
which follows from \eqref{FdefI}
 and the invertibility of $F_d(i\k )$,
one finds the behaviour of the transformation
solution at the origin,
\begin{equation}\label{tsze}
u_c(r\rightarrow 0)= f_d(i\k ,r)[F_d(-i\k )F_d^{-1}(i\k )C+D]+{\rm
o}(r^{\nu})\,.
\end{equation}
We assume
\begin{equation}\label{nonsm}
{\rm det}\left(F_d(-i\k )F_d^{-1}(i\k )C+D\right)\neq 0\,,
\end{equation}
which can always be provided by a
proper choice of matrices $C$ and $D$.
The leading term of the superpotential at
$r\to0$ reads
\begin{equation}\label{wc0}
w_c(r\rightarrow 0)= -r^{-1}\nu +{\rm o}(1)\,,\qquad \nu_j>0\,,
\end{equation}
where we used the Laurent series for the Jost
solution
\begin{equation}\label{se}
f(k,r)=r^{-\nu}(2\nu-1)!!F^T(k)+r^{-\nu+1}b_1(k)+{\rm o}(r^{-\nu+1})\,.
\end{equation}
It follows from \eqref{Vzero} and the \Sc equation that $b_1(k)=0$.
We note the diagonal character of superpotential \eqref{wc0} at the
origin.
The singularity at the origin
of the transformed potential,
\begin{equation}
V_c(r\rightarrow 0)\rightarrow r^{-2}\tilde{\nu}(\tilde{\nu}+1)=
(V_d-2w'_c)|_{r\rightarrow 0}=r^{-2}\nu(\nu-1),
\end{equation}
decreases by one unit, $\nu\rightarrow \tilde{\nu}=\nu-1$.
Hence we can apply our coupling
transformation to potentials for which matrix $\nu$ is
positive definite, $\nu>0$, a property we will assume
to hold throughout the paper.

The Jost matrix can be obtained from expression \eqref{rajs}
of the regular solution $\varphi_c(k,r)$
corresponding to $V_c$.
The regular solution of the transformed potential $\varphi_c(k,r)$
is determined by \eqref{regs} with the singularity parameter $\tilde{\nu}$.
To derive it, we act on both sides of expression \eqref{rajs}
of the regular solution $\varphi_d(k,r)$
for potential $V_d$ with the transformation operator $L_c$.
From \eqref{regs}, \eqref{psit} and \eqref{wc0}, it follows that
\begin{equation}\label{new}
L_c\varphi_d(k,r)=-\varphi_c(k,r).
\end{equation}
Taking into account \eqref{jsc}, we rewrite \eqref{new} as
\begin{equation}\label{fckr1}
\fl \varphi_c(k,r)=
-\frac{i}{2k}\left[f_c(-k,r)(w_{c\infty}+ikI_N)F_d(k)-f_c(k,r)(w_{c\infty}-ikI_N)F_d(-k)\right]
\,.
\end{equation}
Comparing \eqref{rajs} and \eqref{fckr1} we find a relation
between the initial and transformed Jost matrices
\begin{equation}\label{tjm}
F_c(k)  = -(ikI_N+w_{c\infty})F_d(k)\,.
\end{equation}

 For $M=0$ and $M=N$, $Q$ is absent and $A=I_N$ in \eqref{asi}.
When $M=N$,
the superpotential at infinity \eqref{asi} becomes proportional to the
identity matrix, $w_{c\infty}=\k I_N$.
The transformed Jost matrix \eqref{tjm} becomes diagonal.
Similarly, the case $M=0$ leads to $w_{c\infty}=-\k I_N$.
From here we draw an important conclusion.
The necessary (but not sufficient) condition for a
non-trivial coupling in the Jost and hence scattering matrices is
$0<M<N$.
This will be assumed in the following.

As already mentioned,
a non-trivial coupling in the Jost matrix requires not only
a non-diagonal Jost matrix,
but also the impossibility to diagonalize this matrix by a
$k$-independent transformation. It is clear that matrix
$ikI_N+w_{c\infty}$ from \eqref{tjm} can be diagonalized by a
$k$-independent transformation. When $F_d$ is not proportional to
the identity matrix,
 channels in $F_c(k)$ are coupled in a non-trivial way.
 Nevertheless, this property does not guarantee the
 non-triviality of the $S$-matrix.
 As it follows from definition \eqref{SdefI},
  the $S$-matrix
 will be trivially coupled when the product
 $F_d(-k)F_d^{-1}(k)$ is proportional to the identity
 matrix,  i.e. when
\begin{eqnarray}\label{rc}
F_{d;j}(k)=|F_{d;j}(k)|{\rm e}^{-i\delta(k)}\,\Rightarrow\,
S_{d;j}(k)=(-1)^{l_j}{\rm e}^{2i\delta(k)}\,,
\nonumber \\
j=1,\ldots,N\,.
\end{eqnarray}
 In particular, if all single-channel potentials have
 the same $S$-matrix,
 i.e.\ if they are phase
 equivalent \cite{baye:87} or isophase
 \cite{samsonov:02}, the $S$-matrix resulting from a SUSY
 transformation keeps being trivially coupled.
 When the initial $S$-matrix is not proportional
 to the identity matrix,
 one may expect
 non-trivially coupled channels.

The analytic expression \eqref{tjm} for the Jost matrix allows us
to study spectral properties of the transformed potential
\eqref{Vt} \cite{pupasov:08a}. The positions of the bound/virtual
states and resonances are defined as the solutions of
${\rm det\,}F_c(k)=0$.
 As it follows from \eqref{tjm} and \eqref{asi}, the Jost-matrix determinant is given by
\begin{equation}\label{jdz}
{\rm det\,}F_c(k)=(-1)^N(ik+\k)^{M}(ik-\k)^{N-M}{\rm det\,}F_d(k)
\end{equation}
since $w_{c\infty}$ has the $M$ fold
degenerate eigenvalue $\k$ and the $N-M$ fold
degenerate eigenvalue $-\k$.
Therefore, if ${\rm det\,}F_d(k)$ has no pole at $k=\pm i\k$
(this property is assumed to hold in the rest of the paper), the
SUSY transformation leads to a new $M$ fold degenerate bound state with
$k_b=i\k$, $E_b=-\k^2$ and an $N-M$ fold degenerate virtual state with
$k_v=-i\k$, $E_v=-\k^2$.

Now we continue to compare our method with the approach
developed by Amado et al \cite{amado:88b}.
For that, we calculate
the asymptotic behaviour of matrix
$\Phi(r)=[u_c(r)^{\dagger}]^{-1}$, which upon using \eqref{ucas}
reads
\begin{equation}\label{gs}
\Phi(r\to\infty)\to \left(A^T\right)^{-1}\left(
\begin{array}{cc}
I_{M}e^{-\k r} & 0\\
0 & I_{N-M}e^{\k r}
\end{array}
\right).
\end{equation}
The $M$ first columns of $\Phi(r)$ are vectors
decreasing at infinity.
According to
\eqref{tsze} and \eqref{se},
$\Phi(r)$ is a regular solution, $\Phi(0)=0$. Therefore
these vectors correspond to
the bound state wave functions of the coupled system
appeared after the SUSY transformation.
This confirms that the energy level of this bound state
is $M$ fold degenerate.
All the
other columns in $\Phi(r)$ correspond to virtual states.
For the particular case $M=1$
this asymptotic form just corresponds to the transformation
function used  by the authors of \cite{amado:88b} for decoupling a coupled
problem.
We thus conclude that their transformation
corresponds to a particular case of our
transformation when realized in the opposite direction.

Another useful remark is that although the superpotential $w_c(r)$
depends on parameters $X_0$, the Jost matrix $F_c(k)$
and, hence, the $S$-matrix are
$X_0$-independent. This means that superpotential $w_c(r)$ leads
to a family of potentials, parameterized by the entries
of $X_0$, having the same scattering properties.

Below we concentrate on
 the two-channel case with equal thresholds and arbitrary partial waves.
The coupling SUSY transformation produces in this case one
bound state and one virtual state.
First we will
analyze the long range behaviour of the transformed
potential.

%%%%%%%%%%%%%%%%%%%%%%%%%%%%%%%%%%%%%%%%%%%%%%%%18.09.08

%%%%%%%%%%%%%%%%%%%%%%%%%%%%%%%% 19.09.08

\section{Long range behaviour of the transformed potential \label{sec:LR}}

In the two-channel case, according to
\eqref{Vinf},
 the initial diagonal potential has the following long-range
behaviour
\begin{equation}\label{vd}
V_d(r\rightarrow\infty)\to\frac{1}{r^2}\left(
\begin{array}{cc}l_1(l_1+1)& 0\\
0 & l_2(l_2+1)\end{array}\right).
\end{equation}
%Below we will
%find the asymptotic behaviour of the superpotential and transformed
%potential.
The Jost solution at large distances
is expressed in terms of
third kind Bessel functions $H_l^{(1)}(z)$,
also called first Hankel functions (see \cite{Beitman} for a definition)
\begin{equation}\label{jsld}
\fl f_d(k,r\rightarrow\infty)=\diag
\left[h_{l_1}(kr),h_{l_2}(kr)\right]\,,\qquad
h_l(z)=i^{l+1}(\pi z/2)^{\frac{1}{2}} H_l^{(1)}(z)\,.
\end{equation}
The recurrence relations for  $h_l(z)$
%\begin{equation}\label{rs}
%h'_l(kr)=ikh_{l-1}(kr)-lh_l(kr)r^{-1}\,,
%\end{equation}
and its asymptotic behaviour
\begin{equation}\label{asyh}
h_l(kr)=e^{ikr}\left(1+\frac{il(l+1)}{2kr}+{\rm
o}(r^{-1})\right)\,,
\end{equation}
follow from those for $H_l^{(1)}(z)$
 \cite{Beitman}.

For the coupling transformation, according to our previous
discussion, we choose transformation function \eqref{tf2} with
matrices
\begin{equation}\label{mcd}
C=\left(
\begin{array}{cc}1&0\\ q &0\end{array}\right),
\qquad D=\left(
\begin{array}{cc}x&-q\\ 0 &1\end{array}\right),
\end{equation}
which contain only two independent parameters $x$ and $q$. The restriction
on the parameters
\begin{equation}
\qquad x+F_{d,1}(-i\k)F_{d;1}^{-1}(i\k)-q^2F_{d,2}(-i\k)F_{d;2}^{-1}(i\k)\neq 0
\end{equation}
follows from \eqref{nonsm}.
Transformation solution \eqref{tf2} reads
\begin{equation}\label{u2ld}
u_c(r)=\left(
\begin{array}{cc}f_{d;1}(-i\k r)+x f_{d;1}(i\k r)
&-qf_{d;1}(i\k r)\\
qf_{d;2}(-i\k r) &f_{d;2}(i\k r)
\end{array}\right).
\end{equation}

Let us consider the first two terms in the asymptotic behaviour of the
superpotential \eqref{U},
$w_c(r\rightarrow\infty)=w_{c\infty}+w_{-1}r^{-1}+{\rm
o}(r^{-1})$. The first term $w_{c\infty}$ has been calculated for
an arbitrary number of channels in Section \ref{sec:MT}. Thus from
\eqref{asi} we obtain
\begin{equation}\label{w2i}
w_{c\infty} =\frac{\k}{1+q^2} \left(
\begin{array}{cc}1-q^2&2q\\ 2q&q^2-1\end{array}
\right).
\end{equation}
Another parametrization for $w_{c\infty}$ is useful,
\begin{equation}\label{w2ia}
w_{c\infty}=\k\left(
\begin{array}{cc}\cos \alpha& \sin \alpha\\
\sin\alpha & -\cos\alpha\end{array}\right),
\qquad q=\tan\frac{\alpha}{2}\,.
\end{equation}
Note that a non-zero value of $w_{-1}$
will lead to a modification of the long range behaviour of
potential \eqref{Vt}  with $w'_c(r\rightarrow\infty)=
-w_{-1}r^{-2}+{\rm o}(r^{-2}) $.

In order to establish the asymptotic behaviour of the potential
$V_c(r) = V_d(r)-2w_c'(r)$, we replace $f_{d,j}(\pm i\k
r)$ in \eqref{u2ld} by its asymptotic form % $h_{l_j}(\pm i\k r)$,
given in \eqref{jsld}  and
neglect in \eqref{Vt} and \eqref{U}  all exponentially decreasing
terms such as $h_{l_1}(i\k r)h_{l_2}(i\k r)$.
Taking into account
 \begin{equation}
\fl h_{l_1}(i\k r)h_{l_2}(-i\k r)=1+\frac{1}{2\k r}\left[l_1(l_1+1)-
l_2(l_2+1)\right]+{\rm o}(r^{-1})\,,\quad
r\rightarrow\infty,
\end{equation}
combining \eqref{Vt} with \eqref{vd}
and using parametrization \eqref{w2ia}, one finally gets
\begin{eqnarray}\label{vM}
V_{c}(r\rightarrow\infty)=
\frac{1}{r^2} \left(
\begin{array}{cc} l_1(l_1+1) & 0\\
0 & l_2(l_2+1) \end{array}
\right)
\nonumber \\
+
\frac{\left[l_2(l_2+1)-l_1(l_1+1)\right]\sin\alpha}{r^2}
\left(
\begin{array}{cc}
\sin \alpha& -\cos \alpha\\ -\cos\alpha & -\sin\alpha\end{array}\right).
\end{eqnarray}
A similar asymptotic behaviour of the matrix
potential is obtained from the Gelfand-Levitan equation in
\cite{fulton:55}.

From \eqref{vM} we conclude that, for $l_1\neq l_2$, the transformed potential
has a non-zero long range coupling, $V_{c;12}\ne0$.
Moreover, it is impossible
to associate diagonal entries of $V_c$ given in \eqref{vM}
with usual centrifugal terms.
The only way to avoid this inconvenience is to fix $\alpha=\pm\pi/2$
(or equivalently $q=\pm 1$),
which leads to a physically reasonable long range behaviour
\begin{equation}\label{dvM1}
V_{c}(r\rightarrow\infty,q=1)=\frac{1}{r^2}\left(
\begin{array}{cc}l_2(l_2+1) & 0\\ 0 & l_1(l_1+1)\end{array}\right)\,.
\end{equation}
Having compared \eqref{vd} and \eqref{dvM1} we find the modification of
the corresponding angular momentum quantum numbers under the SUSY
transformation
$l=\diag(l_1,l_2)\rightarrow\tilde{l}=\diag(l_2,l_1)$.
For short, this unusual property of the SUSY transformation will be called the
exchange
of the channels angular momenta. Summarizing,
we get an additional constraint $q=1$ (the dual case $q=-1$
leads to the same transformed potential
except $V_{c;12}\rightarrow -V_{c;12}$) to consider only physically
reasonable
potentials in the case $l_2\neq l_1$.

\section{Transformed Jost and scattering matrices,
 eigenphase shifts and mixing angle \label{sec:ST}}

In the two-channels case, introducing $w_{c\infty}$ as given in
\eqref{w2ia} into \eqref{tjm} provides an explicit relation between
the transformed Jost matrix $F_c(k)$ and the initial
diagonal Jost matrix $F_d(k)$,
\begin{equation}\label{tjm2c}
F_c(k)=-\left(
\begin{array}{cc} ik +\k\cos\alpha& \k\sin\alpha\\
\k\sin\alpha & ik -\k\cos\alpha\end{array}\right)F_d(k)\,.
\end{equation}
From \eqref{jdz}, we obtain
${\rm det\,}F_c=(k^2+\k^2){\rm det\,}F_d$.
The coupling transformation produces one bound state and one virtual
state, in agreement
with the general properties of the transformed
Jost matrix analyzed in Section \ref{sec:MT}.

Once the transformed Jost matrix $F_c(k)$ \eqref{tjm2c} is
found,
the $S$-matrix may be obtained
according to its definition \eqref{SdefI},
where
we have to take into account the change of attribution
of the angular momenta
$l\rightarrow\tilde{l}$ by the
 coupling transformation,
\begin{equation}\label{st}
S_c(k)=e^{i\tilde{l}\frac{\pi}{2}}(-ikI_2+w_{c\infty})
(-1)^{l}S_d(k)(ikI_2+w_{c\infty})^{-1}e^{i\tilde{l}\frac{\pi}{2}}\,.
\end{equation}
The diagonal matrix
\begin{equation}\label{st1}
 S_d(k)=\diag(e^{2i\delta_{d;1}(k)},e^{2i\delta_{d;2}(k)})
\end{equation}
is obtained from the diagonal
Jost matrix $F_d(k)$ before the transformation.
One can see that for the particular case of
 identical partial waves, $l\propto I_2$,
our result \eqref{st} reproduces the corresponding
relation (17a) from \cite{amado:88b}.
For different partial waves however,
the modification of the angular momenta leads to
the appearance  of
 additional phase factors $e^{i\tilde{l}\frac{\pi}{2}}$ and $(-1)^{l}$.

Let us now find the transformed eigenphase shifts $\delta_{c;j}(k)$,
$j=1,2$, and the mixing angle
$\epsilon (k)$.
Since the scattering matrix is unitary and symmetric,
 there exists an
orthogonal transformation $R(k)$ diagonalizing this matrix,
\begin{equation}
R_c^T(k)
\left(
\begin{array}{cc}
S_{c;11}(k) & S_{c;12}(k)
\\ S_{c;12}(k) & S_{c;22}(k)
\end{array}
\right)
R_c(k)=
\left(
\begin{array}{cc}
 e^{2i\delta_{c;1}(k)} & 0
\\ 0 & e^{2i\delta_{c;2}(k)}
 \end{array}
 \right).
\end{equation}
The rotation matrix $R_c(k)$ is parameterized
by a mixing angle $\epsilon=\epsilon(k)$,
\begin{equation}
\qquad R_c(k)=
\left(
\begin{array}{cc}
 \cos\epsilon(k) & \sin\epsilon(k)
\\ -\sin\epsilon(k) &\cos\epsilon(k)
\end{array}
\right),
\end{equation}
which is expressed
in terms of $S$-matrix entries as
\begin{equation}\label{ma}
\tan 2\epsilon(k) = \frac{2 S_{c;12}(k)}{S_{c;22}(k)-S_{c;11}(k)}\,.
\end{equation}

Here one can distinguish three essentially different cases:\\
{\bf (i)} the difference between the angular momenta is odd,
$l_2=(l_1+1)\,({\rm mod}\,2)\,$;\\
{\bf (ii)} the difference between the angular momenta is even,
$l_2\neq l_1$, $l_2=l_1\,({\rm mod}\,2) $;\\
{\bf (iii)} the angular momenta coincide, $l_2=l_1$.

Note that case {\bf (i)}  does not correspond
 to any reduction of the rotationally invariant three-dimensional scattering problem,
since in this case any nontrivial coupling means a parity
breakdown (see e.g.\ \cite{taylor:72}). For the sake of
completeness we will analyze this case also although  the
corresponding system of coupled equations has no direct relation
to a scattering problem. Moreover, we will use the usual
scattering theory terminology in this case also, although from the
point of view of a three-dimensional scattering it bears only a formal
character. In cases {\bf (i)} and {\bf (ii)}, we should put
$\alpha=\pi/2$, which is not necessary in the third case (see \eqref{vM}).

 Definition \eqref{st1} allows writing
\begin{equation}
\fl S_{d;1}(k)+S_{d;2}(k)\equiv e^{2i\delta_{d;1}(k)}+e^{2i\delta_{d;2}(k)}=
2 {\rm e}^{i(\delta_{d;1}+\delta_{d;2})} \cos\Delta\,,\qquad \Delta=\delta_{d;2}-
\delta_{d;1}\,,
\end{equation}
and \eqref{ma} leads to expressions
for the mixing angle in the three cases as
\begin{equation}\label{mab1}\fl
{\bf (i)}:\,\tan 2\epsilon(k) = \frac{2
(-1)^{(l_2-l_1-1)/2}\k\sin\alpha\left(k \sin\Delta
-\k\cos\alpha\cos\Delta \right)}{2\k k \cos\alpha\cos\Delta
-(k^2-\k^2)\sin\Delta}\,,
\end{equation}
and
\begin{equation}\label{mab2}\fl
{\bf (ii)}, {\bf(iii)}:\, \tan 2\epsilon(k) = \frac{2(-1)^{(l_2-l_1)/2}
\k\sin\alpha\left(\k\cos\alpha\sin\Delta+k\cos\Delta
\right)}{\sin\Delta\left(k^2-\k^2\cos 2\alpha\right) -2\k k
\cos\alpha\cos\Delta }\,.
\end{equation}
Since $q=1$ and $\alpha=\pi/2$ in cases {\bf (i)} and {\bf (ii)},
expressions \eqref{mab1} and \eqref{mab2}
are simplified to
\begin{equation}\label{mpsp}\fl
{\bf (i)}:\, \epsilon(k) =(-1)^{(l_2-l_1-1)/2}\arctan\frac{k}{\k}\,,
\end{equation}
\begin{equation}\label{mab2zb}\fl
{\bf (ii)}:\,\tan 2\epsilon(k) =(-1)^{(l_2-l_1)/2}\frac{2 \k k}{k^2 +\k^2}\cot\Delta.
\end{equation}

We will assume below that the scattering matrix
of the transformed potential
satisfies the effective range expansion
(see e.g.\ \cite{delves:58}), which implies
\begin{eqnarray}\label{ere}
\cot \delta_{c;1,2}(k\rightarrow 0)=a_{1,2}k^{-(2l_{1,2}+1)}+{\rm o}\left(k^{-(2l_{1,2}+1)}\right),
\nonumber \\
\epsilon(k\rightarrow 0)=\epsilon_0 k^{|l_2-l_1|}+{\rm o}\left(k^{|l_2-l_1|}\right).
\end{eqnarray}
Since there are rather simple analytical expressions
for the mixing angle, we will analyze  restrictions
on parameters of the SUSY transformation which
follow from the second equation in \eqref{ere}.

In case  {\bf (i)}, \eqref{mpsp}
satisfies the effective range expansion \eqref{ere} when $|l_2-l_1|=1$
and violates \eqref{ere} when $|l_2-l_1|>1$.
The important property of the coupling transformation in case
{\bf (i)} is that
the transformed phase shifts coincide with the initial phase shifts,
i.e.,
\begin{equation}\label{spphases}
R_c^T(k)S_c(k)R_c(k)=\left(
\begin{array}{cc} e^{2i\delta_{d;2}(k)} & 0
\\ 0 & e^{2i\delta_{d;1}(k)} \end{array}\right).
\end{equation}
Therefore, one may separately fit the phase shifts for the $l_1$ and $l_2$ waves
before the coupling transformation.

In case
{\bf (ii)}, the effective range expansion
for mixing angle \eqref{mab2zb} leads to the restriction $\cot\Delta = 0$ or
$\delta_{d,2}(0)-\delta_{d,1}(0)=(n+1/2)\pi$ . According to the Levinson theorem (see e.g.\
\cite{newton:82}) this means that the potential $V_d$ supports a
bound state at zero energy.

Finally, in case {\bf(iii)}
there is no any additional constraint since
$\epsilon(k\rightarrow 0)={\rm const}$.

Having established properties of the transformed
phase shifts and the mixing angle,
we will consider in the next section
 some schematic examples of scattering for the $s-s$, $s-p$ and $s-d$
 coupled channels.

\section{Examples \label{sec:E}}

To illustrate the difference between couplings in potential,
Jost and scattering matrices, we construct in this section
nontrivially coupled potentials having trivially coupled
$S$-matrices and both trivially and non-trivially coupled
Jost matrices. After that we exemplify SUSY transformations
leading to non-trivially coupled $S$-matrices.

\subsection{Coupled potentials with uncoupled $S$-matrices\label{Ap:2}}

Let us consider the 1-channel
 potential expressed in terms of a Wronskian as
\begin{eqnarray}\label{Vb}
\fl V(r;\beta)=-2\frac{d}{dr^2}\ln W\left[\sinh(\k_0r),\sinh(\k_2r),\exp(\k_1r)+\beta\exp(-\k_1r)\right],\\
\fl \k_0<\k_1<\k_2\,,\qquad \beta<-1\,,\nonumber
\end{eqnarray}
which can easily be obtained from the zero potential with the help
of the usual (i.e.\ 1-channel) SUSY transformations. This
potential has one bound state at energy $E=-\k_1^2$ and its Jost function has the form
\begin{equation}\label{F_d_ex}
F(k)=i(k-i\k_1)\left[(k+i\k_0)(k+i\k_2)\right]^{-1}\,.
\end{equation}
All
potentials from the $\beta$-family \eqref{Vb} have the same Jost and scattering
matrices.
Therefore, we can construct a diagonal potential
$V_{d}(r)=\diag\left[V(r;\beta_1),V(r;\beta_2)\right]$
with a two fold degenerate bound state at energy $E=-\k_1^2$.
Both its Jost and scattering matrices are proportional
to the identity matrix
\begin{equation}\label{ex1}
 F_d(k)=F(k) I_2\,,
\qquad S_d(k)=F(-k)F^{-1}(k) I_2\,.
\end{equation}
As a result, the Jost matrix \eqref{tjm} obtained after the coupling
transformation
can be diagonalized by the same $k$-independent
transformation as the superpotential $w_{c\infty}$.
 This just corresponds to a trivial coupling in both Jost
 and scattering matrices.

\begin{figure}%[ht]
\begin{center}
%\begin{minipage}{14cm}
\epsfig{file=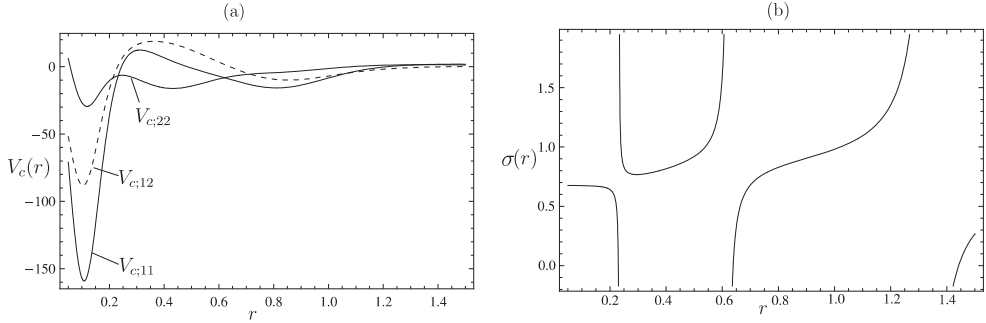, width=13cm}
\caption{\small (a)
Exactly solvable potential $\bar{V}_c=V_c$ obtained from two
uncoupled potentials \eqref{Vb}
($\beta_1=-2\,, \beta_2=-1.5\,, \k_0=1\,, \k_1=2.5\,,\k_2=3.5$)
by the coupling transformation with
parameters $q=0.5$, $\kappa=6$, $x=25$. (b) Ratio
$\sigma(r)=V_{c;12}(r)/\left(V_{c;22}(r)-V_{c;11}(r)\right)$. \label{figSStrpt}}
%\end{minipage}
\end{center}
\end{figure}

For the coupling transformation we choose the transformation function
\eqref{u2ld} where Jost solutions $f_{d}(i\k,r)$ of the \Sc equation
\eqref{se} with potential $V_{d}$ are used. To avoid a
singularity at finite distance in the transformed potential we
impose the restriction $\kappa>\k_2>\k_1$. Such a transformed
potential is shown in figure \ref{figSStrpt}(a). The function
$\sigma(r)=V_{c;12}/(V_{c;22}-V_{c;11})$ demonstrates the
non-triviality of the transformed potential matrix. If
$\sigma$ is a constant, the potential matrix is
 globally diagonalizable.
As we see from figure \ref{figSStrpt} (b),
a non constant $\sigma$ means that the transformed
 potential has a non-trivial coupling.
At the same time, the mixing angle \eqref{mab2} in the scattering matrix
 is just a constant for $\Delta = 0$
\begin{equation}\label{tma}
\epsilon(k)=
-\frac{\a}{2}\,.
\end{equation}
The phase shifts for this potential read
\begin{eqnarray}\label{tcps}
\d_{c;1}(k) & = & 2\pi-\sum\limits_{j=0}^2\arctan\frac{k}{\k_j}+\arctan\frac{k}{\k}\,,\\
\d_{c;2}(k) & = & \pi-\sum\limits_{j=0}^2\arctan\frac{k}{\k_j}-\arctan\frac{k}{\k}\,.
\end{eqnarray}
From here one can see that after the coupling transformation
the additional bound state increases the value of the phase shifts at
zero energy in agreement with the Levinson theorem.

% {b1 -> -2, b2 -> -1.5, k1 -> 1, k2 -> 2.5, x -> 25, q -> 1/2, k3 -> 6,
%  ks -> 3.5}
%\begin{figure}%[ht]
%\begin{center}
%\begin{minipage}{8cm}
%\epsfig{file=sstcps.eps, width=8cm}
%\caption{\small Phase shifts $\delta_{c;1}(k)$, $\delta_{c;2}(k)$ and
%constant mixing parameter $\epsilon$
%for the potential shown in Fig. \ref{figSStrpt}. \label{figsstcps}}
%\end{minipage}
%\end{center}
%\end{figure}

To show the restrictive character of the requirement for the $S$
matrix to be non-trivially coupled, we construct below a potential
with non-trivially coupled potential and Jost matrices
but a trivially coupled $S$-matrix.
This possibility is based on
the fact that in the single-channel case two different Jost
functions may correspond to the same scattering matrix
\cite{sparenberg:96,samsonov:02}.
In this case, the two Jost functions differ from
each other by a real factor for real $k$'s (see \eqref{rc}).
Therefore if we apply our
coupling transformation to the following uncoupled system
\begin{equation}\label{ntc1}
\fl V_d(r)=\diag\left[V_1(r),V_2(r)\right],\ \ \
F_d(k)=\diag\left[F_1(k),F_2(k)\right], \ \ \ S_d(k)=S(k)I_2\,,
\end{equation}
from \eqref{ntc1}, \eqref{Vt}, \eqref{tjm} and \eqref{st} we can
see that the transformed potential and Jost matrices cannot be
diagonalized by a constant rotation whereas the scattering matrix
becomes diagonal after the same $k$-independent rotation which
diagonalizes $w_{c\infty}$.

An example in which we get a non-diagonal Jost matrix and a
trivially coupled $S$-matrix after applying the coupling
transformation follows from \eqref{ntc1} where we choose
%\cite{samsonov:02}
\begin{eqnarray}\label{Ntc1}
F_d(k) & = & \diag\left[\frac{i(k-i\k_0)}{(k+i\k_1)(k+i\k_2)},
\frac{-i}{(k+i\k_0)(k+i\k_1)(k+i\k_2)}\right],\\
S_d(k) & = & \frac{(k+i\k_0)(k+i\k_1)(k+i\k_2)}{(k-i\k_0)(k-i\k_1)(k-i\k_2)}I_2\,.
\end{eqnarray}
Here $\k_0$, $\k_1$ and $\k_2$ are arbitrary real parameters.
Matrix $\nu$ for the corresponding potential
\begin{equation}\label{Vb2}
V_d(r)=\diag\left[V(r,\b <-1),V(r,\b=-1)\right]\,,
\end{equation}
is $\nu=\diag(1,3)$ meaning that $\nu-1>0$ and we can apply the
coupling transformation. Here the non-trivially coupled
transformed Jost matrix \eqref{tjm} with $F_d$ as given in
\eqref{Ntc1} leads to the following trivially coupled $S$-matrix
\begin{equation}
S_c(k)=
(ikI_2-w_{c\infty})^2\frac{(k+i\k_0)(k+i\k_1)(k+i\k_2)}{(k-i\k_0)(k-i\k_1)(k-i\k_2)(k^2+\k^2)}\,.
\end{equation}
The corresponding phase shifts
are given by
\eqref{tcps} where $\d_{c;2}\rightarrow \d_{c;2}-\pi$
and the mixing angle is given by \eqref{tma}.

\subsection{ Coupled $s-s$ partial waves\label{sec:Ss1}}
Using our general scheme described in section \ref{sec:ST},
we can
study the behaviour of the phase shifts for the $s-s$ coupled
potential.
Since the angular momenta in both channels coincide,
we have here the
case {\bf (iii)} discussed above.
Parameter $q$ is not fixed from the
long-range behaviour of the potential and the mixing
angle is given by \eqref{mab2}.
The analysis of this expression is
based on the low-energy behaviour of the phase shifts
$\delta_{d;1,2}$ before the coupling transformation
\begin{equation}\label{les}
e^{2i\delta_{d;1,2}}=1-2i a_{1,2}k+{\rm o}(k)\,,
\end{equation}
where $a_1$ and $a_2$ are the scattering lengths for each channel.
Combining \eqref{mab2} and \eqref{les} we get
\begin{equation}\label{l}
\tan 2\epsilon(k\rightarrow0)=-\frac{2
(1+(a_{1}-a_{2})\kappa\cos\alpha) \sin\alpha}{
2\cos\alpha+(a_{1}-a_{2})\kappa\cos 2\alpha} +{\rm o}(k)\,.
\end{equation}
The expansion of the eigenvalues of the transformed scattering matrix
at low energies reads
\begin{equation}\label{slun}\fl
e^{2i\delta_{c;1,2}}=1- ik\left[(a_1+a_2)\pm\sqrt{(a_2-a_1)^2+
4\left(1/\kappa-(a_2-a_1)\cos\alpha\right)/\kappa}\right] +{\rm
o}(k)\,.
\end{equation}

An important result from the point of view of inverse
scattering corresponds to a coupling vanishing at low
energies, i.e.\ when $\epsilon(0)=0$.
This leads to an additional link between the parameters,
\begin{equation}\label{const}
\cos\alpha=\frac{1}{(a_{2}-a_{1})\kappa}\,,
\end{equation}
where we have used \eqref{l} and \eqref{ere}.
Hence \eqref{slun} simplifies into
\begin{equation}\label{slc}
e^{2i\delta_{c;1,2}}=1-2ia_{2,1}k+{\rm o}(k)\,.
\end{equation}
In this case, the
scattering lengths for the transformed potential
coincide with the initial scattering lengths $a_1$ and $a_2$.
This property allows us to fit low energy scattering data
for uncoupled channels
thus simplifying essentially the inverse problem.
Let us illustrate this property in a schematic example.

We start from the zero potential with a transformation which
introduces poles at the origin, $\nu=\diag (0,0)\rightarrow
\nu=\diag (1,1)$. In each channel we realize the usual
(i.e.\ $1$-channel) SUSY transformation with transformation
functions $\sh(\k_1r)$ and $\sh(\k_2r)$.
This leads to the uncoupled
superpotential
\begin{equation}\label{w1}
w_d(r)=\diag\left[\k_1\coth(\k_1r),\k_2\coth(\k_2r)\right]
\end{equation}
and the potential  (see \eqref{Vt})
\begin{equation}\label{V1}
V_d(r)=2\diag\left[\k_1^2\csh^2(\k_1r),\k_2^2\csh^2(\k_2r)\right]\,,
\end{equation}
with the Jost solution
\begin{equation}\label{f1}\fl
f_d(k,r)=e^{ikr} \left(
\begin{array}{cc}k+i\k_1\coth(\k_1r)&0\\
0&k+i\k_2\coth(\k_2r)\end{array}\right) \left(
\begin{array}{cc}k+i\k_1&0\\
0&k+i\k_2\end{array}\right)^{-1}
\end{equation}
 and the Jost matrix
\begin{equation}
F_d(k)=(w_{d\infty}-ikI_2)^{-1}\,,\quad
w_{d\infty}=\lim_{r\to\infty}w_d(r)= {\rm diag}\left(\k_1,
\k_2\right).
\end{equation}
As coupling transformation we choose transformation function
\eqref{tf2} with matrices $C$ and $D$ given by \eqref{mcd}. The
explicit expression for $u_c(r)$ coincides with \eqref{u2ld} where
\begin{equation}\label{g12}
\fl f_{d;1}(i\k,r)=\frac{\k+\k_1\coth(\k_1r)}{\k+\k_1}\,e^{-\k r}\,,
\qquad
 f_{d;2}(i\k,r)=\frac{\k+\k_2\coth(\k_2
r)}{\k+\k_2}\,e^{-\k r}\,.
\end{equation}
The parameter $x$ from \eqref{mcd}
 should be chosen in order to avoid any
singularity
in the transformed potential.
As can easily be seen from the
analysis of ${\rm det\,}u_c$,
it is sufficient to choose $x$
large enough.
The asymptotic behaviour of the
superpotential is given by \eqref{w2i} or \eqref{w2ia}.

The Jost matrix $F_c(k)$ may be found from \eqref{tjm}. Its explicit
expression is rather involved and we omit it. More important is
its determinant \eqref{jdz}, the expression of which is extremely simple,
\begin{equation}\label{detf2}
{\rm det\,}\,F_c(k)= \frac{k^2+\k^2}{(k+i\k_1)(k+i\k_2)}\,.
\end{equation}
From here we find the location of the bound state at $k_b=i\k$ and
the virtual state at $k_v=-i\k$.

%%%%%%%%%%%%%%%%%%%%%%%%%%%%%%%%%%%%%%%%%%%%%%%%%%%
% Amado discussion is moved to the end of the file
%%%%%%%%%%%%%%%%%%%%%%%%%%%%%%%%%%%%%%%%%%%%%%%%%%%

\begin{figure}%[ht]
\begin{center}
\begin{minipage}{10cm}
\epsfig{file=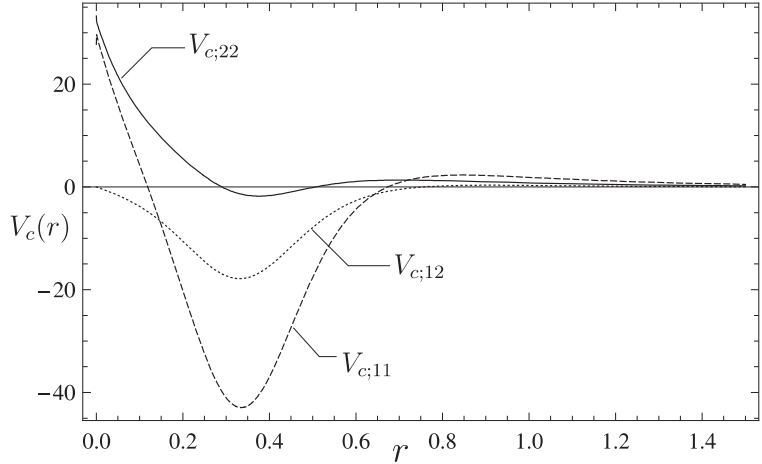, width=10cm} \caption{\small Exactly
solvable $s-s$ potential $\bar{V}_c=V_c$ with parameters
$\k_1=1.5$, $\k_2=1$, $q=0.4$, $\kappa=4.14286$, $x=15$.
\label{figVss}}
\end{minipage}
\end{center}
\end{figure}

\begin{figure}%[ht]
\begin{center}
%\begin{minipage}{15cm}
\epsfig{file=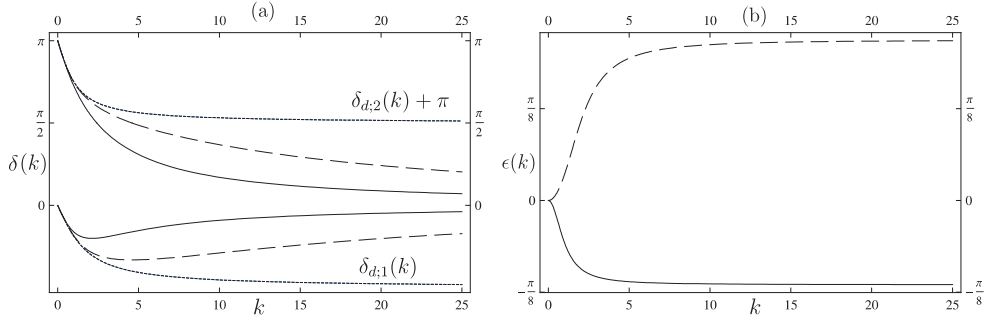, width=13cm} \caption{\small The
scattering matrix for the coupled $s-s$ potential. (a) The
eigenphases are shown as dotted lines for $V_d$ and as dashed and
solid lines for $V_c$. The parameters are: solid lines -
$\k_1=1.5$, $\k_2=1$, $q=0.4$, $\kappa=4.14286$, $x=15$, dashed
lines - $\k_1=1.5$, $\k_2=1$, $q=1.2$, $\kappa=13.6667$, $x=15$.
(b) Mixing angle $\epsilon$. \label{figssps}}
%\end{minipage}
\end{center}
\end{figure}

The chain
of two SUSY transformations with parameters
$\k_{1,2}=a_{1,2}^{-1}$ and $\kappa, q, x$ described above leads
to the mixing angle \eqref{l}.
The corresponding potential ($q=0.4$) is shown in figure \ref{figVss}.
The factorization constant $\k$ is fixed from \eqref{const}.
As a result, the mixing angle takes the form
\begin{equation}\label{mpss}
\tan 2\epsilon(k)=-\frac{2 k^2 \k_1 \k_2 \tan\alpha}
{ \k_1^2 \k_2^2 \sec^2\alpha + k^2 (\k_1^2 + \k_2^2) }\,.
\end{equation}
Parameters $\k_1$ and $\k_2$ are related with $1$-channel transformations
and allow us to fit the
scattering lengths.
The mixing angle $\epsilon(k)$
depends on parameter $\alpha$,
 which allows one to
fit its experimental behaviour at low energies.
The
mixing angle at large energies tends to a constant value, $\tan
2\epsilon(k\to\infty)=-2\k_1\k_2\tan\alpha/(\k_1^2+\k_2^2)$ which
can also be fitted using corresponding experimental data (if
available).
Figure \ref{figssps} shows the phase shifts and mixing
angle
 for two coupled $s-s$ potentials.

The phase shifts of the diagonal potential $V_d$ are shown as dotted
lines in figure
\ref{figssps}(a).
The phase shifts of the transformed potential
$V_c$ are shown
 as dashed ($q=0.4$) and solid ($q=1.2$) lines
respectively.
One can see that the slopes of these curves coincide
at the origin.
The mixing angles of the transformed potential are plotted in figure
\ref{figssps}(b).

According to \eqref{detf2} this potential
has one bound state at the factorization energy $E_g=-\kappa^2$.
 Note that the normalization constant of the bound state wave
function is determined by parameter $q$ as follows from
\eqref{gs}.

%{s1 -> 1.5, s2 -> 1, b2 -> 0.4, k3 -> 3.25711, b1 -> 15}

\subsection{Coupled $s-p$ partial waves\label{sec:SP}}

\begin{figure}%[ht]
\begin{center}
\begin{minipage}{10cm}
\epsfig{file=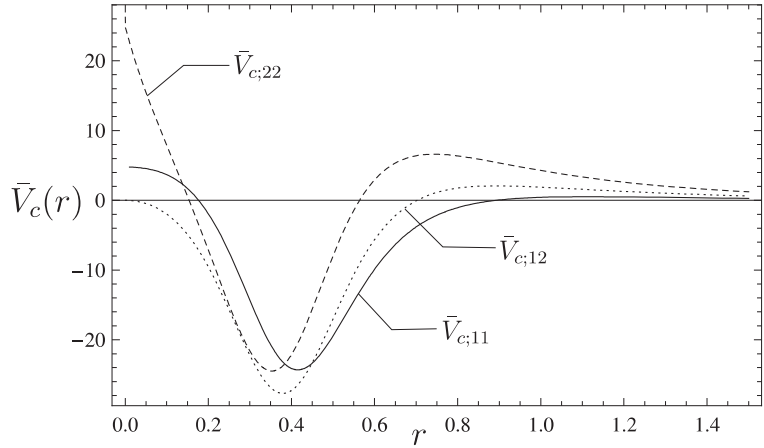, width=10cm} \caption{\small Exactly
solvable $s-p$
potential $\bar{V}_c=V_c-\tilde{l}(\tilde{l}+1)r^{-2}$
with parameters $\k_0=1.5$, $\k_1=1.75$, $\kappa=3.53$, $q=1$,
$x=1$. \label{figVps}}
\end{minipage}
\end{center}
\end{figure}
%\begin{figure}%[ht]
%\begin{center}
%\begin{minipage}{10cm}
%\epsfig{file=de1.eps, width=8cm}
%\caption{\small
%The scattering matrix for the coupled $s-p$ potential.
%The $s$-wave phase shift $\delta_{c;2}=\delta_{d;1}$
%is shown
%as a solid line.
%The $p$-wave phase shift $\delta_{c;1}=\delta_{d;2}=0$
%is not presented.
%The mixing angle $\epsilon$ is shown as a dashed line.
%the parameters are
%$\k_0=1.5$, $\k_1=1.75$, $\kappa=3.53$.
%\label{figSPps}}
%\end{minipage}
%\end{center}
%\end{figure}

In this section we consider the simplest $s-p$ coupled potential.
This potential is characterized by $\tilde{l}=\tilde{\nu}=\diag
(1,0)$. The coupling transformation acts as follows:
\begin{equation}\fl
l=\diag(l_1,l_2)\rightarrow\tilde{l}=\diag(l_2,l_1)\,,\qquad
\nu=\diag(\nu_1,\nu_2)\rightarrow\tilde{\nu}=\diag(\nu_1-1,\nu_2-1)\,.
\end{equation}
Therefore the initial diagonal potential should have
$l=\diag (0,1)$, $\nu=(2,1)$.
These properties are satisfied for the initial potential of
the form
\begin{equation}\label{ipsp}
V_d(r)=\diag\left[
-2\frac{d^2}{dr^2}
\ln W[\sinh\left(\k_0r\right),\sinh\left(\k_1r\right)],2r^{-2}\right],
\end{equation}
where the potential in the $s$-channel is obtained from
the zero potential after two consecutive
SUSY transformations with $\k_0$ and $\k_1$ as factorization constants.
The potential in the $p$-channel
is just the centrifugal term.
The Jost solution in the $s$-channel is expressed in terms of
the
Wronskian of factorization solutions $\sinh(\k_jr)$,
$j=0,1$,
\begin{equation}\label{jssp}
\fl f_d(k,r)=\diag\left[
\frac{W\left[\sinh\left(\k_0r\right), \sinh\left(\k_1r\right),
e^{ikr}\right]}{(k+i\k_0)(k+i\k_1)W\left[\sinh\left(\k_0r\right),
 \sinh\left(\k_1r\right)\right]},\frac{(i+kr){\rm e}^{ikr}}{kr} \right].
\end{equation}
The uncoupled Jost matrix reads
\begin{equation}\label{fspi}
F_d(k)=\diag\left[
-\left[(k+i\k_0)(k+i\k_1)\right]^{-1}, ik^{-1}\right].
\end{equation}

The next step is to apply the coupling transformation with
the transformation function \eqref{tf2} where the Jost solution
is replaced by
\eqref{jssp}.
An example of potential curves is shown in figure
\ref{figVps}.
The Jost matrix \eqref{fspi} is transformed according to
 \eqref{tjm} and the scattering matrix is given by \eqref{st}.

Since the transformed eigenphase shifts coincide
with the initial phase shifts,
we may fit the phase shifts for the $s$ and $p$ waves
separately before the coupling transformation.
In our example the phase shifts read
\begin{equation}
\delta_{c;1}(k)=0\,,
\qquad \delta_{c;2}(k)=\pi-\arctan\frac{k}{\k_0}-\arctan\frac{k}{\k_1}\,.
\end{equation}
Parameters $\k_0$ and $\k_1$ allow one to fit the $s$-channel phase shifts.
Parameter $\kappa$ may be used
to fit the slope of the mixing angle \eqref{mpsp}
at zero energy.
If necessary, one may use arbitrary chains of 1-channel
transformations
to get the best fit of the phase shifts.

\subsection{Coupled $s-d$ partial waves\label{sec:SD}}

The simplest $s-d$ coupled potential is characterized by
$\tilde{l}=\tilde{\nu}=\diag (2,0)$. Therefore the initial diagonal
potential should have $l=\diag (0,2)$ and $\nu=(3,1)$.
Moreover, as
we established in section \ref{sec:ST},
$\delta_{d;2}(0)-\delta_{d;1}(0)=(n+1/2)\pi$ which
for $n=0$
leads to the following initial phase shifts $\delta_{d;2}(0)=\pi/2$ and
$\delta_{d;1}(0)=0$.

We start with the initial $s$-wave potential
\begin{equation}
V_{0}(r)=\frac{-2\k_0^2}{\cosh^2(\k_0r)}
\end{equation}
having a zero energy virtual
state \cite{newton:82}
which follows from its Jost function
\begin{equation}
F_{0}(k)=\frac{k}{k+i\k_0}\,.
\end{equation}
Note that this potential and the
solutions of the corresponding \Sc equation may be obtained by
a SUSY transformation.
This is a regular potential.
To be able to apply the coupling transformation, we
increase its singularity
at  the origin using three SUSY transformations with the
 transformation functions
\begin{equation}\label{urki}
u(\k_i,r)=\k_i \sinh(\k_ir)+\k_0\cosh(\k_ir)\tanh(\k_0r)\,,\qquad i=1,2,3\,.
\end{equation}
The potential and the Jost function in the $s$-channel
after these transformations read
\begin{eqnarray}
V_{d;1}(r)=\frac{-2\k_0^2}{\cosh^2(\k_0r)}-
2\left(\ln W\left[u(\k_1,r),u(\k_2,r),u(\k_3,r)\right]\right)'',
\\
F_{d;1}(k)=\frac{-ik}{(k+i\k_0)(k+i\k_1)(k+i\k_2)(k+i\k_3)}\,.
\end{eqnarray}
The potential in the $d$-channel
\begin{eqnarray}
\fl V_{d;2}(r) & = & \frac{6}{r^2}-2 \ln v''(\k_4,r)
=\frac{6(3 + 6\k_4x+6\k_4^2r^2+4\k_4^3r^3+\k_4^4r^4)}
{r^2(3+3\k_4r+\k_4^2r^2)^2}\,,
\\
\fl v(\k_4,r) & = & {\rm e}^{-\k_4r}\left(1+{3}{\k_4r}+\frac{3}{(\k_4r)^2}\right),
\end{eqnarray}
is obtained from the centrifugal term $6/r^2$ by the SUSY transformation
with $v(r,\k_4)$ as the transformation function,
which
decreases the singularity of the potential at the origin.

\begin{figure}%[ht]
\begin{center}
%\begin{minipage}{8cm}
\epsfig{file=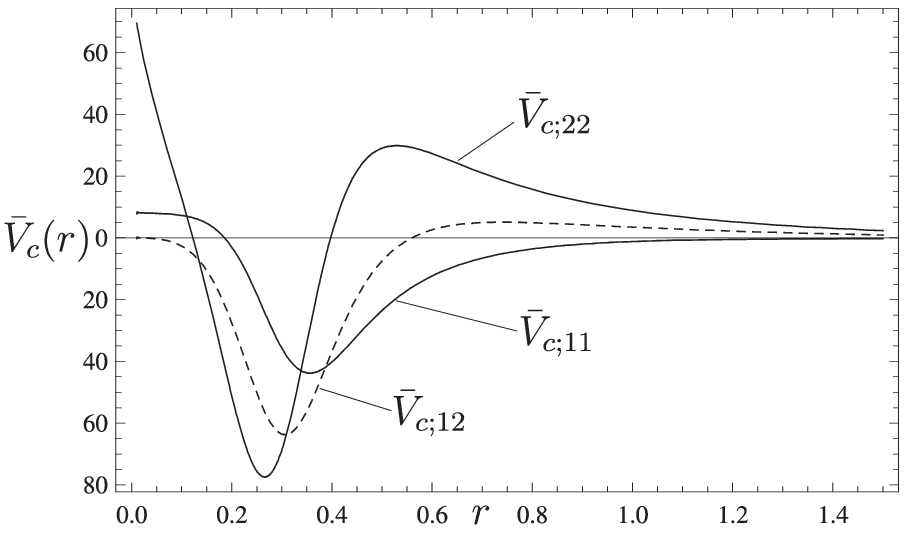, width=8cm} \caption{\small Exactly solvable
$s-d$ potential $\bar{V}_c=V_c-\tilde{l}(\tilde{l}+1)r^{-2}$ with
parameters $\k_0=1$, $\k_1=1.5$, $\k_2=1.75$, $\k_3=2$,
$\kappa_4=3$ $q=1$, $x=15$, $\k=5.53$. \label{figVds}}
%\end{minipage}
\end{center}
\end{figure}
\begin{figure}%[ht]
\begin{center}
%\begin{minipage}{15cm}
\epsfig{file=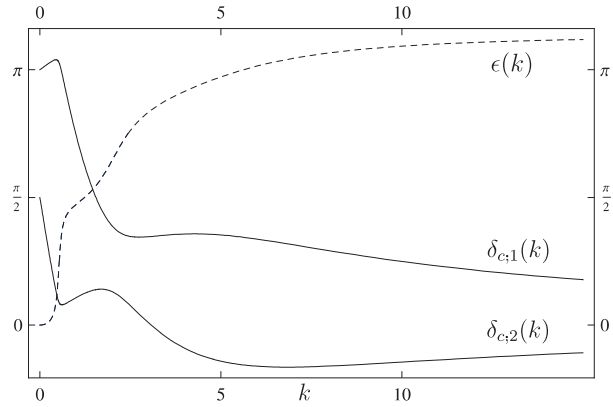, width=8cm} \caption{\small The
scattering matrix for the coupled $s-d$ potential. The phase
shifts $\delta_{c;1}(k)$ and $\delta_{c;2}(k)$ are plotted by
solid lines. The mixing angle $\epsilon(k)$ is plotted by the
dashed line. The corresponding parameters are $\k_0=1$,
$\k_1=1.5$, $\k_2=1.75$, $\k_3=2$, $\kappa_4=3$ $q=1$, $x=15$,
$\k=5.53$. \label{figSDps}}
%\end{minipage}
\end{center}
\end{figure}

The Jost solution in the
$s$-channel is expressed in terms of the Wronskian of
factorization solutions \eqref{urki}
\begin{eqnarray}\label{Js1}\fl
f_{d;1}(k,r) & = &
\frac{W\left[u(\k_1,r),u(\k_2,r),u(\k_3,r),
f_0(k,r)\right]}{(k+i\k_1)(k+i\k_2)(k+i\k_3)W\left[u(\k_1,r),u(\k_2,r),u(\k_3,r)
\right]}\,,
\\
\fl f_0(k,r) & = &
{\rm e}^{ikr}\frac{k+i\k_0\tanh(\k_0r)}{k+i\k_0}\,.
\end{eqnarray}
The Jost solution in the $d$-channel is
\begin{equation}\label{jd1}\fl
f_{d;2}(k,r)=i
\frac{W\left[v(\k_4,r),
h_2(kr)\right]}{(k-i\k_4)v(\k_4,r)}\,,
\qquad
h_2(kr)={\rm e}^{ikr}\left(1+{3i}{\k_4r}-\frac{3}{(\k_4r)^2}\right).
\end{equation}
The uncoupled Jost matrix reads
\begin{equation}\label{fsdi}
F_d(k)=\diag\left[\frac{-ik}{(k+i\k_0)(k+i\k_1)(k+i\k_2)(k+i\k_3)}, \frac{ik-\k_4}{k^2}\right],
\end{equation}
which produces the eigenphase shifts
\begin{equation}\label{pssd}
\delta_{d;1}(k)=\frac{\pi}{2}-\sum\limits_{j=0}^{3}\arctan\frac{k}{\k_j}\,,\qquad
\delta_{d;2}(k)=\arctan\frac{k}{\k_4}\,.
\end{equation}

Next we apply the coupling transformation with the
transformation function \eqref{tf2} where the Jost solution
$f_d=\diag(f_{d;1},f_{d;2})$
is combined from \eqref{Js1} and \eqref{jd1}.
An example of potential curves thus obtained is shown
in figure \ref{figVds}.
The Jost matrix \eqref{fsdi} is
transformed according to
 \eqref{tjm} and the scattering matrix is given by \eqref{st}.

The corresponding phase shifts and mixing angle are plotted in
figure \ref{figSDps}. The mixing parameter $\epsilon(k)$ is
determined by \eqref{mab2zb} which, in the current case, reduces
to
\begin{equation}\label{mpsd}
\tan 2\epsilon(k) = \frac{2 \k k}{(k^2
+\k^2)}\tan\left(\sum\limits_{j=0}^{4}\arctan\frac{k}{\k_j}\right).
\end{equation}
We were not able to find simple expressions
 for the eigenphase shifts in this case.
One can see that the mixing parameter satisfies
 the effective range expansion \eqref{ere}
(see \eqref{mpsd} and figure \ref{figSDps}).
Unfortunately, this
 is not the case for the phase shifts (see figure \ref{figSDps}).

\section{Conclusion \label{Concl}}

In the present paper,
a careful analysis
of SUSY transformations between
 uncoupled potentials with equal
 thresholds
but arbitrary partial waves in $N$ channels
and coupled ones is given.
In particular, we formulated conditions imposed on
the transformation function
to
get a nontrivially coupled scattering matrix.
A family of iso-phase potentials
 generated by a coupling SUSY
transformation has been obtained.
This family is parameterized
by a real symmetric $M\times M$, $0<M<N$, non-singular matrix $X_0$.
The analysis of the zeros
of the Jost-matrix determinant for these potentials
showed that the
SUSY transformation creates a new $M$ fold degenerate
bound state energy $E_b=-\k^2$ and an $N-M$ fold degenerate virtual state
energy $E_v=-\k^2$.

In the most important practical case,
the two-channel case,
 we analyzed the behaviour
of the superpotential and potential at large distances in details.
We have found an unusual effect, i.e.\
a modification of the long-range behaviour of the potential
under a coupling SUSY transformation
which consists in
 the exchange of attribution of the partial waves between the channels.
Moreover, we analyzed in details
the transformed scattering matrix, phase shifts
and mixing angle.
The analysis of the
phase shifts and mixing angle demonstrated how
scattering properties change
after a SUSY transformation.

As an illustration of our approach, several simple examples have been presented.
First, to emphasize the difference between couplings in the potential,
Jost and scattering matrices,
we presented examples of a trivially coupled scattering matrix corresponding to
non trivially coupled potential and Jost matrices.
These examples answer the general questions raised in the introduction:
situations may exist where a non trivially coupled potential leads to a trivially coupled
$S$-matrix, with either a trivially or non trivially coupled Jost matrix.
Thus, the requirement that an $S$-matrix is
non trivially coupled is more restrictive than
the similar requirement for a potential matrix.
Moreover, even the requirement that the Jost matrix is
non trivially coupled is less restrictive than the
corresponding requirement for the $S$-matrix.

Afterwards, a non trivial coupling was introduced in the $s-s$, $s-p$
and $s-d$ channels.
In both $s-s$ and $s-p$ examples,
we have shown how to fit the low-energy behaviour of the phase shifts
and mixing angle
using parameters of the transformation.
We note that the $s-p$ example contains all essential ingredients
of a convenient inversion scheme since the $s-p$ coupling transformation
conserves the initial phase shifts,
which allows separating the inversion of phase shifts from that of coupling parameters.
In the $s-d$ case,
 to satisfy the effective range expansion for the mixing parameter,
we
used an initial potential with a zero energy virtual state.
Nevertheless,
the obtained phase shifts of the coupled $s-d$ potential
do not satisfy the correct effective range expansion.
 Moreover, the
presence of the zero energy virtual state
strongly restricts possible applications of our method to the inversion
in this case.
We believe that a chain of two coupling transformations
may allow avoiding this inconvenience.

\section*{Acknowledgements}

This work is partially supported by the grants RFBR-09-02-00009a
and SS-871.2008.2.
BFS thanks the National Fund for Scientific Research, Belgium,
for support during his stay in Brussels.
This text presents research results of the Belgian program P6/23
on interuniversity attraction poles of the
Belgian Federal Science Policy Office.

\end{document}